\documentclass[prd,amsmath,amssymb,superscriptaddress,preprintnumbers,twocolumn,10pt]{revtex4-1}

\usepackage{graphicx}
\usepackage{dcolumn}
\usepackage{bm}
\usepackage{amssymb}
\usepackage{latexsym}
\usepackage{booktabs}
\usepackage{amsmath}
\usepackage{multirow}
\usepackage{url}
\usepackage{footnote}

\usepackage{float}
\usepackage[colorlinks=true, linkcolor=red, citecolor=blue]{hyperref}

\usepackage[normalem]{ulem}
\usepackage{color}
\usepackage{array}
\usepackage{enumerate}

\begin{document}


\title{Efficient parameter inference for gravitational wave signals in the presence of transient noises using temporal and time-spectral fusion normalizing flow}

\author{Tian-Yang Sun}
\affiliation{Key Laboratory of Cosmology and Astrophysics (Liaoning) \& College of Sciences, Northeastern University, Shenyang 110819, China}
\author{Chun-Yu Xiong}
\affiliation{Key Laboratory of Cosmology and Astrophysics (Liaoning) \& College of Sciences, Northeastern University, Shenyang 110819, China}
\author{Shang-Jie Jin}
\affiliation{Key Laboratory of Cosmology and Astrophysics (Liaoning) \& College of Sciences, Northeastern University, Shenyang 110819, China}
\author{Yu-Xin Wang}
\affiliation{Key Laboratory of Cosmology and Astrophysics (Liaoning) \& College of Sciences, Northeastern University, Shenyang 110819, China}
\author{Jing-Fei Zhang}
\affiliation{Key Laboratory of Cosmology and Astrophysics (Liaoning) \& College of Sciences, Northeastern University, Shenyang 110819, China}
\author{Xin Zhang}\thanks{Corresponding author}
\email{zhangxin@mail.neu.edu.cn}
\affiliation{Key Laboratory of Cosmology and Astrophysics (Liaoning) \& College of Sciences, Northeastern University, Shenyang 110819, China}
\affiliation{Key Laboratory of Data Analytics and Optimization for Smart Industry (Ministry of Education), Northeastern University, Shenyang 110819, China}
\affiliation{National Frontiers Science Center for Industrial Intelligence and Systems Optimization, Northeastern University, Shenyang 110819, China}

\begin{abstract}
Glitches represent a category of non-Gaussian and transient noise that frequently intersects with gravitational wave (GW) signals, thereby exerting a notable impact on the processing of GW data. The inference of GW parameters, crucial for GW astronomy research, is particularly susceptible to such interference. In this study, we pioneer the utilization of a temporal and time-spectral fusion normalizing flow for likelihood-free inference of GW parameters, seamlessly integrating the high temporal resolution of the time domain with the frequency separation characteristics of both time and frequency domains. Remarkably, our findings indicate that the accuracy of this inference method is comparable to that of traditional non-glitch sampling techniques. Furthermore, our approach exhibits a greater efficiency, boasting processing times on the order of milliseconds. In conclusion, the application of a normalizing flow emerges as pivotal in handling GW signals affected by transient noises, offering a promising avenue for enhancing the field of GW astronomy research.

\end{abstract}
\maketitle

\section{Introduction}

Since the initial detection of gravitational waves (GWs) stemming from binary black hole (BBH) mergers \cite{LIGOScientific:2016aoc} and a binary neutron star (BNS) event \cite{LIGOScientific:2017vwq,LIGOScientific:2017zic,LIGOScientific:2017ync}, the LIGO-Virgo-KAGRA collaboration \cite{LIGOScientific:2014pky,VIRGO:2014yos,KAGRA:2018plz} has reported the detections of over 90 GW events involving compact binary coalescences. These GW observations play a pivotal role in advancing fundamental physics \cite{LIGOScientific:2016lio,LIGOScientific:2018dkp,LIGOScientific:2020tif,LIGOScientific:2021sio,Isi:2019aib,Yunes:2016jcc}, astrophysics \cite{LIGOScientific:2018cki,Annala:2017llu,Margalit:2017dij,LIGOScientific:2016vpg,Mandel:2021smh,Broekgaarden:2021efa}, and cosmology \cite{LIGOScientific:2017adf,Bird:2016dcv,Sakstein:2017xjx,Wang:2018lun,Zhang:2018byx,Zhang:2019ylr,Zhang:2019ple,Wang:2019tto,Zhang:2019loq,Zhao:2019gyk,Jin:2020hmc,Wang:2021srv,Jin:2021pcv,Song:2022siz,Li:2023gtu,Jin:2023sfc,Barack:2018yly,Palmese:2021mjm}. However, it is important to note that GW data are susceptible to non-Gaussian noise contamination \cite{LIGOScientific:2018mvr,LIGOScientific:2020ibl,KAGRA:2021vkt,LIGOScientific:2017tza}, which potentially influences the aforementioned scientific analyses.

Non-Gaussian noise, referred to as glitches and non-stationary noise, poses a challenge due to it contaminating GW data, thereby impacting the quest for GW signals and the accurate estimation of wave source parameters \cite{Macas:2022afm,Abbott_2020,Canton_2014,LIGOScientific:2017tza,LIGO:2021ppb,Powell:2018csz}. Various sources contribute to glitches, including seismic events such as earthquakes \cite{Biscans_2018} and environmental factors like passing trains \cite{Glanzer:2023hzf}. While numerous techniques are presently employed to mitigate glitches at their source  \cite{aLIGO:2020wna,AdvLIGO:2021oxw,Schwartz_2020,LIGOScientific:2021kro}, future observations operating at heightened sensitivities are anticipated to yield increased detection rates \cite{Punturo:2010zz,Dominik:2014yma,LIGOScientific:2016wof}. Consequently, the reduction of GW signals tainted by glitches becomes a challenging task \cite{Soni:2021cjy}.

During the recent third LIGO-Virgo observing run (O3), 24\% of the GW candidates were found to be contaminated by glitches \cite{LIGOScientific:2020ibl,KAGRA:2021vkt}. Notably, the analyses of prominent events like GW170817, GW200129, and GW191109\_010717 were substantially impacted due to glitches \cite{Pankow:2018qpo,Hannam:2021pit,Payne:2022spz,LIGOScientific:2021sio}. Therefore, effective glitch mitigation strategies are imperative prior to undertaking parameter estimation.

The two primary methods commonly employed for complete glitch removal are BayesWave \cite{Cornish:2014kda,Cornish:2020dwh} and gwsubtract \cite{Davis:2018yrz,Davis:2022ird}. Additionally, various approaches exist for mitigating the effects of glitches in observational data \cite{Cornish:2021wxy,Powell:2018csz,Steltner:2021qjy,Talbot:2021igi,LIGO:2020zwl,Hourihane:2022doe,Mohanty:2023mjn,Udall:2022vkv}. However, as of O3, there remains a lack of reasonably low-latency methods for data cleaning \cite{Macas:2022afm}, posing a potential hindrance to the discovery of certain physical phenomena, such as the observation of subsequent electromagnetic (EM) counterparts following BNS mergers. This challenge arises from the computational intensity of the full Bayesian approach, making it a time-consuming process. Consequently, there is a pressing need to develop precise and low-latency deglitching methods.

One prevalent technique for alleviating the impact of corrupted glitches is gating \cite{Usman:2015kfa}, enabling the prompt removal of glitch-affected data with low latency \cite{Messick:2016aqy}. A notable case is the handling of GW170817, for which contaminated segments of the data were excised to facilitate the search for EM counterparts \cite{Pankow:2018qpo}. However, the confluence of gating and signals in the time-frequency domain poses challenges to parameter inference \cite{Pankow:2018qpo}. Machine learning emerges as a promising solution for addressing this issue, given its non-linear GPU-based computational capabilities, making it well-suited for the low-latency processing of non-stationary data \cite{jordan2015machine}. In addition, the robustness of machine learning also makes it more suitable for processing GW data contaminated by glitches \cite{Wei:2019zlc,Ren:2022hny,Jin:2023ahl,Chatterjee:2021lit}.

Several studies have explored the application of machine learning in reconstructing glitches from data, enabling subsequent subtraction to mitigate their effects \cite{Biswas:2013wfa,Yu:2021swq,Vajente:2019ycy,Ormiston:2020ele,Merritt:2021xwh,10.1093/mnras/stad341,Bini:2023gil}. Following the acquisition of clean data, additional computations are often necessary for deriving source parameters for subsequent analyses \cite{Mogushi:2021cpw}. Consequently, we aim to address the query of whether machine learning can be effectively employed to directly infer GW parameters from contaminated data.

Neural posterior estimation \cite{rezende2015variational,NIPS2016_6aca9700,lueckmann2017flexible,greenberg2019automatic}, relying on a normalizing flow, demonstrates a precise estimation of the posterior distribution of source parameters \cite{Dax:2021tsq,Gabbard:2019rde,Chua:2019wwt}. Functioning as a likelihood-free method, a normalizing flow proves effective in handling non-Gaussian data, exemplified by its successful application to the $21$ cm signal \cite{hassan2022hiflow,Zhao:2021ddh,Zhao:2022ren,Zhao:2023tep}. Consequently, our exploration aims to ascertain the viability of employing a normalizing flow in the processing of GW data contaminated by glitches.

In this study, we introduce a novel method grounded in a normalizing flow for parameter estimation for data afflicted by glitches. While utilizing time-frequency domain data proves advantageous for glitch-contaminated data \cite{George:2018awu,9666903,GEORGE201864}, inherent limitations in time-frequency resolution and binning may result in the loss of intricate details, thereby influencing parameter inference. {In recent years, some researchers have studied improvements of the network performance from the data fusion of time series and corresponding frequency domain data \cite{yang2022unsupervised,liu2023temporal,zhang2023co}.} Therefore we employ a dual approach, incorporating both time-domain and time-frequency domain data in the parameter inference process, i.e., temporal and time-spectral fusion normalizing flow (TTSF-NF). Our investigation specifically targets high signal-to-noise ratio (SNR) glitches that defy resolution through robustness.

The organization of this work is as follows: Section~\ref{sec2} provides an introduction to the methodology employed. In Section~\ref{sec3}, we comprehensively present the results yielded by our approach. The conclusion is encapsulated in Section~\ref{sec4}.

\section{Methodology}\label{sec2}
\subsection{Data generation}\label{sec2.1}

\begin{figure*}[!htp]
\centering
\includegraphics[width=1.0\textwidth]{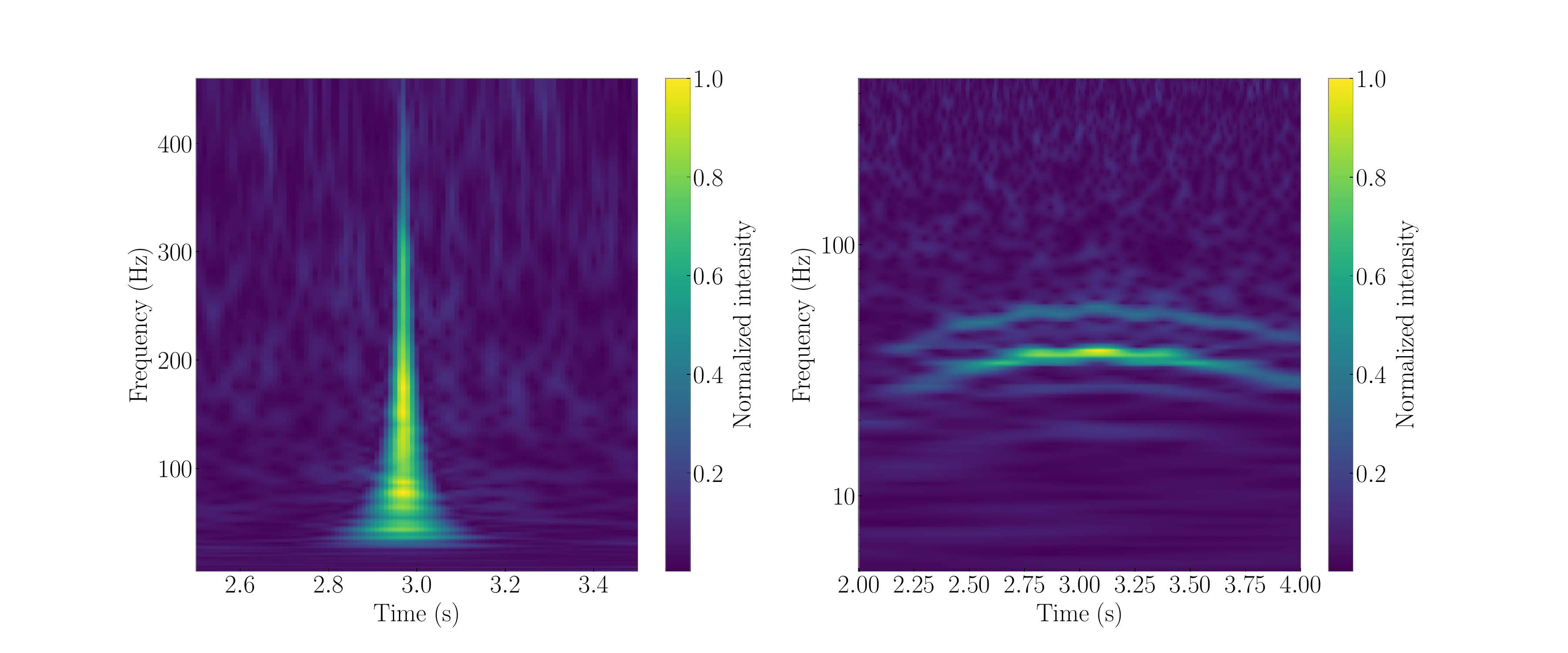}
\centering \caption{\label{fig1} Two types of glitches considered in this work. Left panel: Blip noise characterized by durations on the order of milliseconds and a broad frequency bandwidth on the order of 100 Hz. Right panel: scattered light noise which persists for an extended duration and exhibits a frequency below 100 Hz.}
\end{figure*}

\begin{figure}[!htp]
\centering
\includegraphics[width=0.5\textwidth]{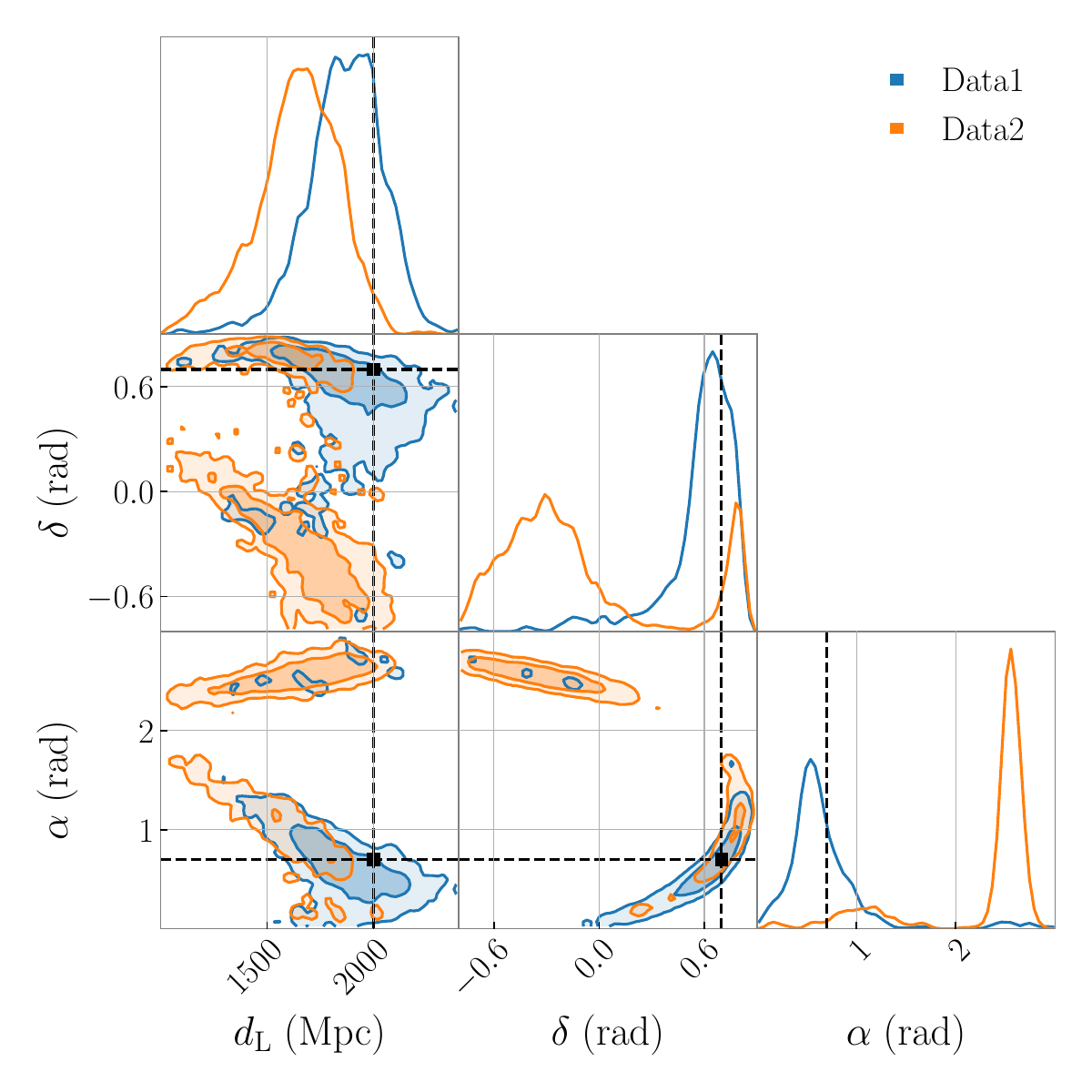}
\centering \caption{\label{fig2} One- and two-dimensional marginalized posterior distributions for $d_{\rm L}$, $\delta$, and $\alpha$ using Data1 and Data2. The intersection points of the dashed lines are the injected parameters. Note that Data1 are the posterior parameters predicted by T-NF with Gaussian noise plus the GW signal and Data2 are the posterior parameters predicted by T-NF with Gaussian noise with blip plus the GW signal.}
\end{figure}

In this study, we focused on two prevalent glitches commonly observed in aLIGO detectors--- namely, ``blip'' and ``scattered light''--- both sourced from Gravity Spy \cite{coughlin2021gravity}. The spectrogram diagrams for each glitch are illustrated in Fig.~\ref{fig1}. Our glitch selection specifically targeted segments with an SNR exceeding 12, a criterion chosen due to the prevalent focus of existing research on scenarios with a relatively low SNR \cite{Powell:2018csz,Du:2023plr}. As depicted in Fig.~\ref{fig2}, relying solely on the robustness of the time normalizing flow (T-NF) proves insufficient for resolving this challenge. Notably, in the case of Hanford's blip during O3b, the proportion of data with $\mathrm{SNR}>12$ is 54\%. This underscores the likelihood of encountering simultaneous occurrences of a signal and a high SNR glitch in future observations.


Blip glitches exhibit a brief duration ($\sim 0.1$~ms) and a frequency ranging from tens of Hz to hundreds of Hz, precisely aligning with the frequency range of BBH \cite{Cabero:2019orq}. Their shape easily resembles that of GW signals \cite{Nitz:2017lco,Cabero:2019orq}. The origin of blip glitches is presently unknown, and they do not manifest in any auxiliary channels \cite{Nuttall:2015dqa}. Consequently, it is probable that they will continue to be a significant component of glitches in the future.


Scattered light glitches typically exhibit persistence and a low frequency. This type of noise arises when a fraction of the laser beam light scatters due to defects in the detector mirror or excessive ground motion, subsequently recombining with the main beam \cite{Accadia:2010zzb,Glanzer:2023hzf}. Although its occurrence diminishes with advancements in the manufacturing processes \cite{KAGRA:2021vkt}, it tends to escalate with increased sensitivity. Consequently, achieving complete elimination in the future is likely to be challenging.

\begin{figure*}[!htp]
\centering
\includegraphics[width=1\textwidth]{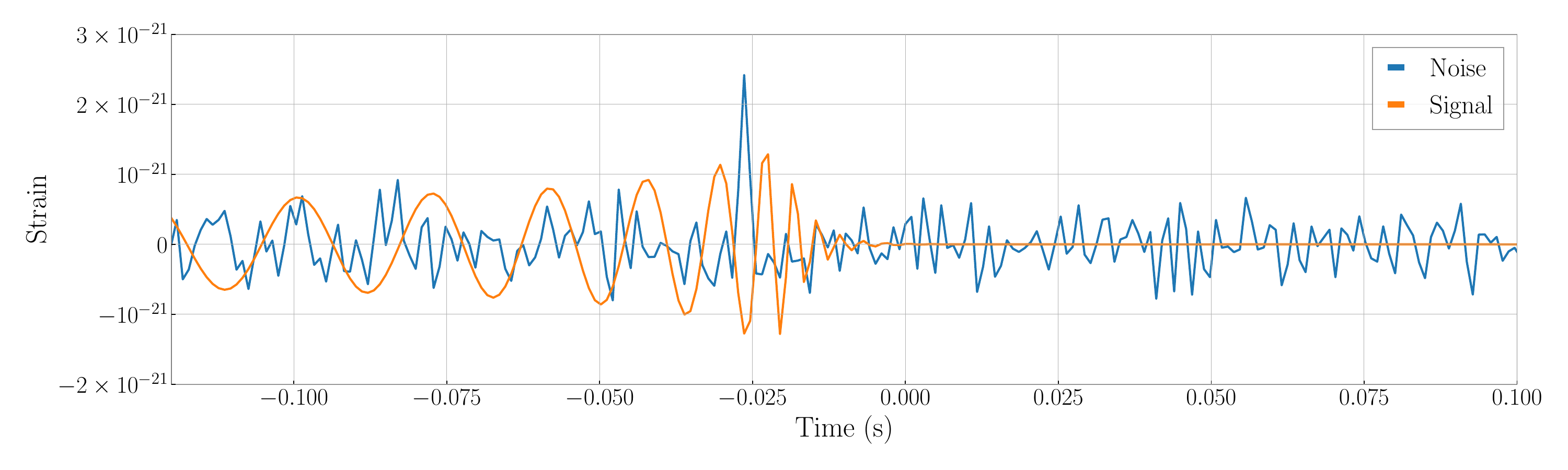}
\centering \caption{\label{fig3} Time-domain strains of noise and GW signal. Here, the noise strain data is adopted from LIGO Hanford noise and the GW signal is generated based on the GW150914-like event. Note that the blip is injected at $t_0 - 30$ ms of the GW signal (this is a situation depicted in Figure 3 of Ref.~\cite{Macas:2022afm}).}
\end{figure*}

In our approach, we utilized data with a duration of 2 s. Considering the higher accuracy in determining the merger time during signal search \cite{Gebhard:2019ldz,Nousi:2022dwh}, we held the GPS time of the merger relative to the center of the Earth fixed at $t_0=1.8$ s and then projected it onto the detector based on the corresponding parameters. Leveraging the work of Alvarez-Lopez \emph{et al.} \cite{Alvarez-Lopez:2023dmv}, who introduced a model capable of concurrently identifying glitches and GW signals in the data, we conducted separate training for each glitch type. To replicate the coincidence of the glitch and the merger moment, we injected glitches at a temporal distance of ($-0.1$ s, $+0.1$ s) from the merger moment. The time strain data can be written as
\begin{equation}
s(t)=h(t)+n(t),
\end{equation}
where $h(t)$ is the GW signal and $n(t)$ is the background noise (including the glitch). The specific time strain is shown in Fig.~\ref{fig3}.


Referring to Hourihane \emph{et al.} \cite{Hourihane:2022doe}, glitches have a more pronounced impact on high-quality systems compared with on low-quality systems. Consequently, this study primarily focuses on higher mass BBH systems. Several representative parameters were considered, and the expectation is that our TTSF-NF will demonstrate applicability to other parameter sets as well. Specific parameters are detailed in Table~\ref{tab1}, where $f_{\rm max}<512$ Hz, and events with $8<\mathrm{SNR}<32$ were exclusively selected. Given the substantial amount of existing observational data, we employed real Hanford noise containing glitches, calibrated according to Gravity Spy. The noise in other detectors corresponds to the glitch-free segments obtained from Gravitational Wave Open Science Center (GWOSC) \cite{trovato2019gwosc}. Adhering to the Nyquist sampling law, the data's frequency only needs to exceed 2 times the value of $f_{\rm max}$; hence, we downsampled the data to 1024 Hz. We introduced the GW signal into the noise, subsequently whitening it to derive the simulated time-domain data.


The spectrogram utilized in this study was generated through the $Q$-transform \cite{Chatterji:2004qg}, a modification of the short Fourier transform. It maps the time-frequency plane with pixels of constant $Q$ (quality factor) by employing an analysis window whose duration is inversely proportional to the frequency. Recognizing the necessity for real-time data analysis, the spectrogram's resolution is constrained to maintain efficiency. To accommodate TTSF-NF's requirements, time and frequency were divided into 200 bins to achieve the necessary image size. The observational data $d$ input to TTSF-NF consists of both time strain data and the spectrogram. All these steps were implemented using $\tt PyCBC$ \cite{Biwer:2018osg}, with the BBH waveform modeled using IMRPhenomPv2 \cite{Khan:2018fmp}.


\begin{table}
\caption{Distribution of simulated GW waveform parameters. Note that other parameters not mentioned are set to zero for simplicity.}\label{tab1}
\centering
\setlength\tabcolsep{18pt}
\renewcommand{\arraystretch}{1.5}
\begin{tabular}{cc}
\hline \hline Parameter & Uniform distribution \\
\hline Chirp mass & $\mathcal{M}_{\rm c}\in[25.0,62.5]~ M_{\odot}$ \\
Mass ratio & $q\in [0.5,1]$\\
Right ascension & $\alpha\in[0, 2\pi]~{\rm rad}$ \\
Declination & $\delta\in[-\pi/2,\pi/2]~{\rm rad}$ \\
Polarization angle & $\psi\in[0,2 \pi]~{\rm rad}$ \\
Luminosity distance & $d_{\rm L}\in[300,3000]~\mathrm{Mpc}$\\
\hline \hline
\end{tabular}
\end{table}

\subsection{Normalizing flow}

The backbone of TTSF-NF is based on the normalizing flow (NF) \cite{rezende2015variational}, a generative model. The fundamental concept of NF involves providing a reversible transformation $f_d$ for each observation data $d$, thereby converting the simple base distribution into a more intricate posterior distribution. The pivotal aspect of this network lies in the reversibility of the transformation and the straightforward computation of the Jacobian matrix. Currently, NF has found extensive application in GW signal processing \cite{Green:2020dnx,Dax:2021tsq,Williams:2021qyt,Langendorff:2022fzq,Ruhe:2022ddi,Ruan:2023fce,Du:2023plr}.


The NF can be expressed by the following formula \cite{rezende2015variational}
\begin{equation}\label{eq1}
q(\theta \mspace{-5mu} \mid \mspace{-5mu} d)=\pi\left(f_d^{-1}(\theta)\right)\left|\operatorname{det} J_{f_d}^{-1}\right|.
\end{equation}

The basic distribution $\pi(u)$ can, in principle, be arbitrary. However, for ease of sampling and density evaluation, it is often chosen to be a standard multivariate normal distribution with the same dimension $D$ as the sample space. {In practice, Eq.~(\ref{eq1}) can be continuously applied to construct arbitrarily complex densities \cite{rezende2015variational}
\begin{equation}
f_d(u)=f_{d,N}\circ f_{d,N-1}\circ \dots \circ f_{d,1}(u),
\end{equation}
where $f_{d,i}(u)~(i=1, \dots ,N)$ represents a block of the NF.} This approach is referred to as neural posterior estimation (NPE). The objective of NPE is to train a parameter conditional distribution that approximates the true posterior distribution. This task translates into an optimization problem, with the aim of aiming to minimizing the expected Kullback-Leibler (KL) divergence \cite{10.1214/aop/1176996454} between these two distributions.


The loss of NF can be written as the expected value (over $d$) of the cross entropy between the true and model distributions \cite{papamakarios2019sequential}
\begin{equation}
L=\mathbb{E}_{p(d)}\left[\operatorname{KL}\left(p(\theta \mspace{-5mu} \mid \mspace{-5mu} d) \| q(\theta \mspace{-5mu} \mid \mspace{-5mu} d)\right)\right].
\end{equation}
On a minibatch of training data of size $N$ , we approximate \cite{Green:2020dnx}
\begin{equation}
L \approx-\frac{1}{N} \sum_{i=1}^N \log q\left(\theta^{(i)} \mspace{-5mu} \mid \mspace{-5mu} d^{(i)}\right).
\end{equation}

It is evident that the training of NF necessitates only the parameters corresponding to each data point without making any assumptions. This constitutes a likelihood-free method, eliminating the need for data modeling. Hence, it is well-suited for non-Gaussian and other challenging model scenarios.

{In this study, we employed a currently more potent flow known as the neural spline flow (NSF) \cite{durkan2019neural}. The fundamental concept behind NSF revolves around a fully differentiable module founded on monotonic rational-quadratic splines. Specifically, it consists of a series of coupling transforms. For $n$-dimensional data, the coupling transform maps the input vector $x$ (for the $(i+1)$th block,  $x=f_{d,i}\circ \dots \circ f_{d,1}(u)$}) to the output $y$ in the following way \cite{Green:2020dnx,Ruan:2023fce}.

1. Divide the input $x$ into two parts, $x = \left [x_{1:m}, x_{m+1:n}\right ]$ where $m<n$.

{2. Input $x_{1:m}$ and the results output by the preceding neural network based on the input data into another neural network (specifically, the residual neural network), to obtain the vector $\alpha$.}

{3. For each $\alpha_{i}$ ($i = m,... , n$), one can construct the invertible function $g_{\alpha_ {i}}$ for computing $y_{i}= g_{\alpha _{i}} (x_{i})$.}

4. Set $y_{1:m} = x_{1:m}$.

{Finally, return $y = [y_{1:m}, y_{m+1:n}]$. The neural spline coupling transform, meanwhile, treats each output as a monotonically increasing segmented function, and by putting the abovementioned $\alpha_ {i}$ through a series of operations one can obtain the knots $\left \{ \left ( u_{i}^{(k)},c_{i}^{(k)} \right ) \right \} _{k=0}^{K} $ and the positive-valued derivatives $\left \{ \delta_{i}^{(k)} \right \} _{k=0}^{K} $, which can be expressed in terms of interpolating rational quadratic (RQ) functions; these RQ samples are differentiable and have analytic inverses, so that they satisfy the properties required for coupled transformations. The NSF directly controls knots and derivatives by adjusting the residual network.}



\subsection{Network architecture}

In the ongoing fourth LIGO-Virgo observing run (O4), a total of four detectors are in operation. Given the temporal correlation between signals from different detectors, the data dimension is substantial, leading to some redundancy. Consequently, a prevalent approach for data processing (feature extraction) involves the utilization of multi-input machine learning \cite{Dax:2021tsq}. In this context, we employed the front-end residual net (ResNet) to extract features from the data $d$.


In this section, we outline the architecture of the ResNet-50 \cite{he2016deep} and employed the normalizing flow model. Two distinct networks, namely 1D ResNet-50 and 2D ResNet-50, were utilized to handle the input time strain and its spectrogram, respectively. Their outputs were then combined into a unified 1D vector.

ResNet-50, a 50-layer deep network, incorporates a unique feature known as skip connections. These connections link the output of one layer to another by bypassing intermediate layers. Combined with these skip connections, the layers form residual blocks, enhancing the appropriateness of the initial weights by enabling the network to learn residuals, with the output typically approaching 0. The activation function for each residual block is rectified linear units (ReLU), effectively capturing complex representations and addressing the vanishing gradient problem.

An innovative aspect of our approach involves the addition of a dropout layer with a value of 0.2 after each activation function within the ResNet. This strategic inclusion mitigates the risk of overfitting and diminishes the network's reliance on input features, potentially enhancing its suitability for processing data containing glitches.


For the subsequent normalizing flow model, we employed the ReLU activation function with the hidden layer sizes set to 4096, 9 flow steps, 7 transform blocks, and 8 bins.

\begin{figure*}[!htp]
\centering
\includegraphics[width=1.0\textwidth]{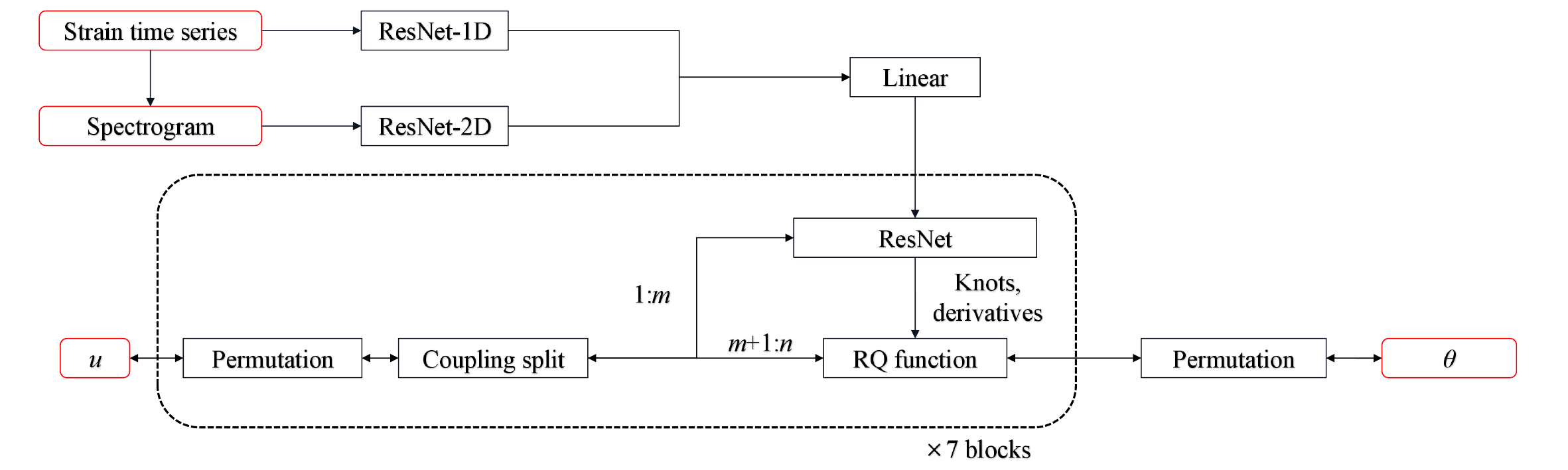}
\centering \caption{\label{fig4} The workflow of TTSF-NF. The input data comprises 2-second time strain data and a spectrogram. The data traverse through the different ResNet-50 network constructed using 1D convolutional layers and 2D convolutional layers, respectively. The features extracted from the two ResNet-50 models are subsequently merged into a 1D feature vector. This feature vector is then employed as a conditional input for the normalizing flow, generating samples from the base distribution and transforming them into the posterior distribution.}
\end{figure*}

In this study, our focus centered on testing events resembling GW150914, which is a representative event. Consequently, we constructed a two-detector network, although the network's applicability extends to scenarios with four or more detectors. The specific structure of the front-end ResNet is detailed in Tables \ref{tab2} and \ref{tab3}, while the network's specific architecture is illustrated in Fig.~\ref{fig4}. In cases in which only the time strain or spectrogram was input for comparison, the corresponding network was omitted accordingly. The network employed in this study is implemented using {\tt PyTorch} \cite{paszke2019pytorch} and {\tt Lampe} \cite{lampe}. Figures were generated using {\tt Matplotlib} \cite{hunter2007matplotlib} and {\tt ligo.skymap} \cite{ligo_skymap}.



Throughout the training process, we iteratively generated data by sampling the prior distributions of the events and obtaining a new noise realization for each data frame. This approach serves to augment the number of training sets and minimizes the risk of NPE producing overly confident or excessively conservative posterior distributions \cite{hermans2022trust}. The AdamW optimizer \cite{loshchilov2018fixing} was employed with a learning rate set to 0.0001, a batch size of 200, and a learning rate decay factor of 0.99. For initializing the network parameters with initial random values, the ``Xavier'' initialization \cite{kumar2017weight} was applied. This initialization method aims to maintain an appropriate scale for the weights during both forward and backward propagation, mitigating issues such as gradient disappearance or explosion. Its design principle involves initializing weights to random values that adhere to a specific distribution, ensuring consistency in the input and output variances.

Training incorporated an early stopping strategy to achieve convergence and prevent overfitting. The network underwent approximately 16 days of training on a single NVIDIA GeForce RTX A6000 GPU with 48 GB of memory.


\begin{table}
\caption{The architecture of the ResNet model for time-domain strain data in the two-detector network. The first column shows the name of the used layer. The second column shows the dimensions of the output data obtained by the corresponding layer. The third column shows the specific structure of each layer, with labels $a$ and $b$ denoting the one-dimensional residual block. Here, each filter size is denoted by $a$, and the output channel is represented by $b$. ``stride $c$'' means that the stride of the convolution or pooling layer is $c$.}\label{tab2}
\centering
\setlength\tabcolsep{11pt}
\renewcommand{\arraystretch}{1.5}
\begin{tabular}{ccc}
\hline \hline Layer name & Output size & Architecture \\
\hline Input layer & (2048, 2) & - \\
Conv1 & (1024, 64) & Cov1D(7, 64), stride 2 \\
MaxPooling & (512, 64) & 3 max pool, stride 2 \\
Conv2\_x & (256, 256) & $\begin{bmatrix}
 1,64\\
 3,64\\
 1,256
\end{bmatrix}\times 3$ \\
Conv3\_x & (128, 512) & $\begin{bmatrix}
 1,128\\
 3,128\\
 1,512
\end{bmatrix}\times 3$ \\
Conv4\_x & (64, 1024) & $\begin{bmatrix}
 1,256\\
 3,256\\
 1,1024
\end{bmatrix}\times 3$ \\
Conv5\_x & (32, 2048) & $\begin{bmatrix}
 1,512\\
 3,512\\
 1,2048
\end{bmatrix}\times 3$ \\
Flatten & (2048) & - \\
\hline \hline
\end{tabular}
\end{table}

\begin{table}
\caption{Same as Table~\ref{tab2}, but the filter size is $a\times a$.}\label{tab3}
\centering
\setlength\tabcolsep{8pt}
\renewcommand{\arraystretch}{1.5}
\begin{tabular}{ccc}
\hline \hline Layer name & Output size & Architecture \\
\hline Input layer & (200, 200, 2) & - \\
Conv1 & (100,100, 64) & Cov2D(7$\times$7, 64), stride 2 \\
MaxPooling & (50, 50, 64) & 3$\times$3 max pool, stride 2 \\
Conv2\_x & (25, 25, 256) & $\begin{bmatrix}
 1\times1,64\\
 3\times3,64\\
 1\times1,256
\end{bmatrix}\times 3$ \\
Conv3\_x & (13, 13, 512) & $\begin{bmatrix}
 1\times1,128\\
 3\times3,128\\
 1\times1,512
\end{bmatrix}\times 3$ \\
Conv4\_x & (7, 7, 1024) & $\begin{bmatrix}
 1\times1,256\\
 3\times3,256\\
 1\times1,1024
\end{bmatrix}\times 3$ \\
Conv5\_x & (4, 4, 2048) & $\begin{bmatrix}
 1\times1,512\\
 3\times3,512\\
 1\times1,2048
\end{bmatrix}\times 3$ \\
Flatten & (2048) & - \\
\hline \hline
\end{tabular}
\end{table}

\section{Results and discussion}\label{sec3}
\subsection{Results and reliability}

\begin{figure}[!htp]
\centering

\includegraphics[width=0.5\textwidth]{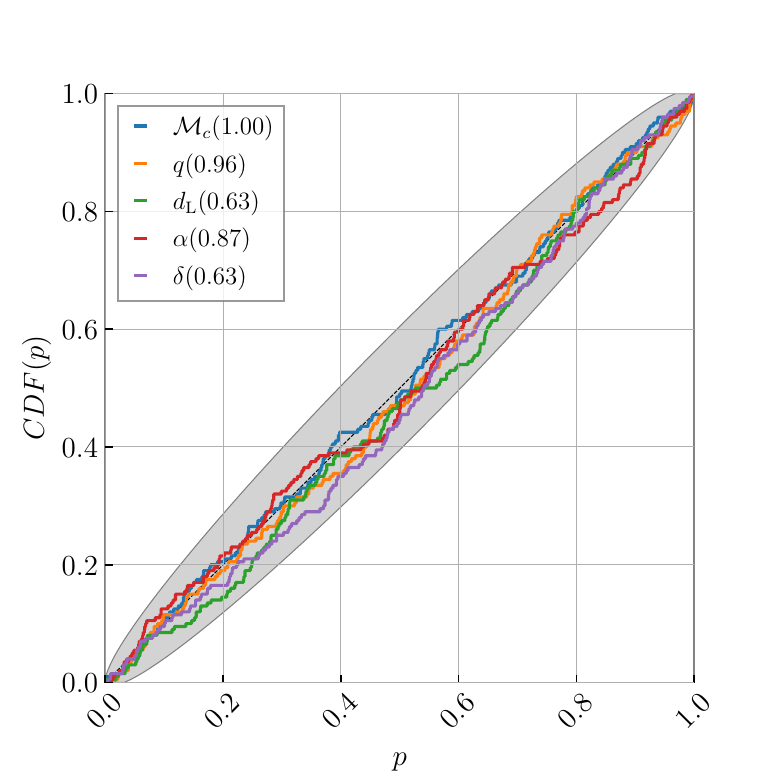}
\centering \caption{\label{fig5} P-P plot for 200 simulated datasets containing blips analyzed by TTSF-NF.} 
\end{figure}

Before drawing specific inferences, it is crucial to assess the reliability of the method's results. This is achieved by conducting the Kolmogorov-Smirnov (KS) test \cite{Veitch:2014wba} to compare the one-dimensional posterior distributions generated by the TTSF-NF outputs. Taking blips as an example, Hanford's O3b data comprises a total of 1437 entries with $\mathrm{SNR}>12$, and 1300 of these were selected for training. Subsequently, KS testing was performed using 100 noise-injected simulated waveforms containing blips, which were not part of the training process. Figure~\ref{fig5} illustrates the construction of a probability-probability (P-P) plot based on 200 simulated datasets. For each parameter, we calculated the percentile value of the true value within its marginal posterior, plotting the cumulative distribution function (CDF) of this value, which represents the range of intervals covered by the corresponding confidence interval. In a well-behaved posterior, the percentiles should be uniformly distributed, resulting in a diagonal CDF. The $p$-values for the KS test are provided in the figure legends, and the gray area denotes the $3\sigma$ confidence limit. The proximity of the CDF to the diagonal indicates the model's ability to accurately sample the posterior, confirming the reasonability of the parameter range given by this method.


\begin{figure}[!htp]
\centering
\includegraphics[width=0.5\textwidth]{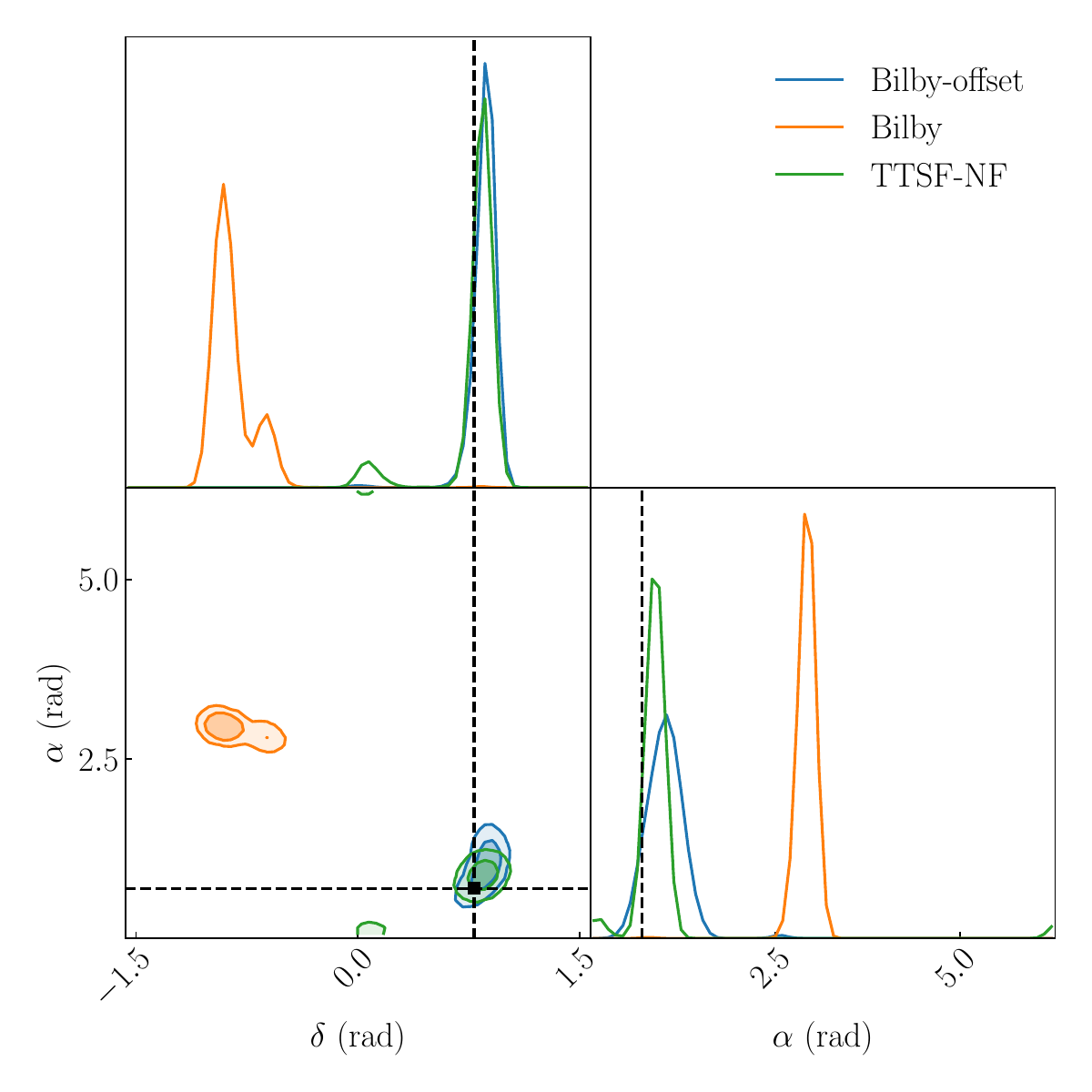}
\centering \caption{\label{fig6} One- and two-dimensional marginalized posterior distributions for $\delta$ and $\alpha$. The intersection points of the dashed lines are the injected parameters. Here, the Bilby-offset represents the posterior parameters predicted by {\tt Bilby} with noise (without blip) plus the GW signal. Bilby represents the posterior parameters predicted by {\tt Bilby} with noise (with blip) plus the GW signal. TTSF-NF represents the posterior parameters predicted by the network with noise (with blip) plus the GW signal.}

\end{figure}

Figure~\ref{fig6} displays the posterior distribution of an event, emphasizing our innovative approach to exploring non-Gaussian noise and proposing new solution ideas. To initially evaluate the effectiveness of this method, we compared it with the posterior distributions in the absence of glitches. Position parameter inference results from {\tt Bilby} \cite{Ashton:2018jfp} and the network are shown for scenarios with and without blips in the noise. The interval time chosen for this comparison, following Macas \emph{et al.} \cite{Macas:2022afm}, is ($t_0 - 30$) ms, when the blip has the most significant impact on positioning. The results indicate that TTSF-NF can avoid position errors induced by glitch-contaminated data and achieves an accuracy comparable to {\tt Bilby} with the {\tt dynesty} sampler \cite{Speagle:2019ivv} on glitch-free data. Although we observe a bimodal shape in the network's posterior distribution, the probability of the second peak is low and can be disregarded. This phenomenon may arise from the narrow width of ResNet and the relatively simple feature extraction \cite{9422007}.


\subsection{Comparative analysis}

\begin{figure}[!htp]
\centering
\includegraphics[width=0.5\textwidth]{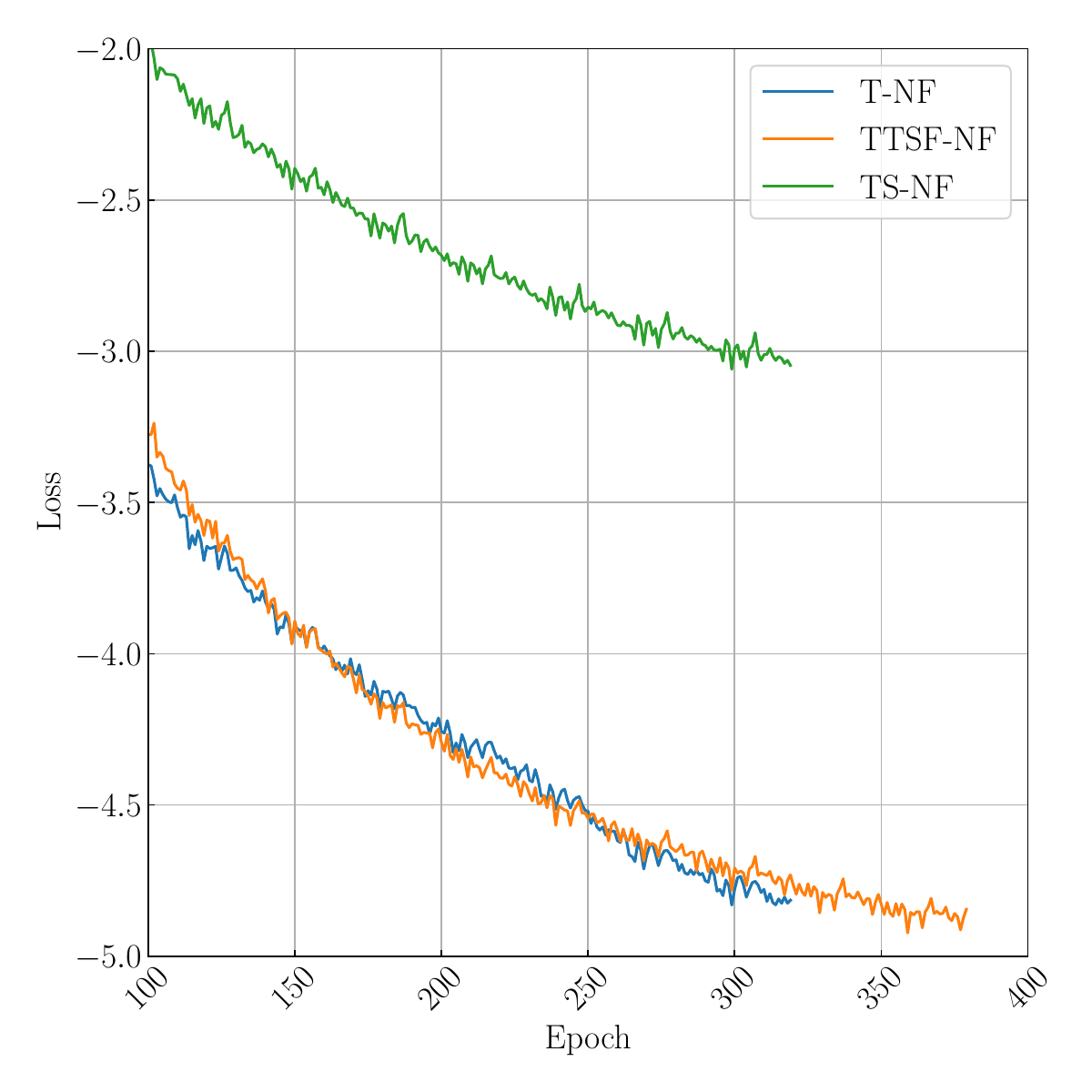}
\centering \caption{\label{fig7} The loss of all different networks for each epoch. {During the training process, we used 10 cycles of early stops to ensure that each network could reach the optimal.} Here, T-NF represents the loss of the network using only time strain. TTSF-NF represents the loss of the network using both time strain and a spectrogram. TF-NF represents the loss of the network using only a spectrogram.}
\end{figure}

We proceeded to investigate whether it is necessary to input both the time strain and the spectrogram. Initially, we examined the overall loss of the network, as depicted in Fig.~\ref{fig7}. The loss comparison suggests that TTSF-NF exhibits a better performance, although T-NF still performs reasonably well. However, these losses are based on the 1300 blips in the training set. To ascertain if the network has genuinely learned the characteristics of blips, we evaluated its generalization performance on the remaining 100 glitches that were not part of the training set.


{For glitches not included in the training set, TTSF-NF outperforms T-NF. The overall parameter estimation performance is quantified by the determination coefficient $R^2$ under the same source parameters, defined as follows:
\begin{equation}
R^{2} =1-\frac{\sum (y_{\text{pre}} -y_{\text{true}})}{\sum (y_{\text{pre}} -\bar{y} _{\text{true}})} ,
\end{equation}
where $y_{\text{pre}}$ and $y_{\text{true}}$ represent the variable estimates in the sample and the true value of the test sample, respectively, and $\bar{y} _{\text{true}}$ represents the average true value of the test sample. We calculated $R^2$ for 200 test datasets in our preset range for both networks. For T-NF, $R^2=0.66$; for TTSF-NF, $R^2=0.71$. Therefore, incorporating both time strain and a spectrogram yields significant improvements.}

\begin{figure*}[!htp]
\centering
\includegraphics[width=0.5\textwidth]{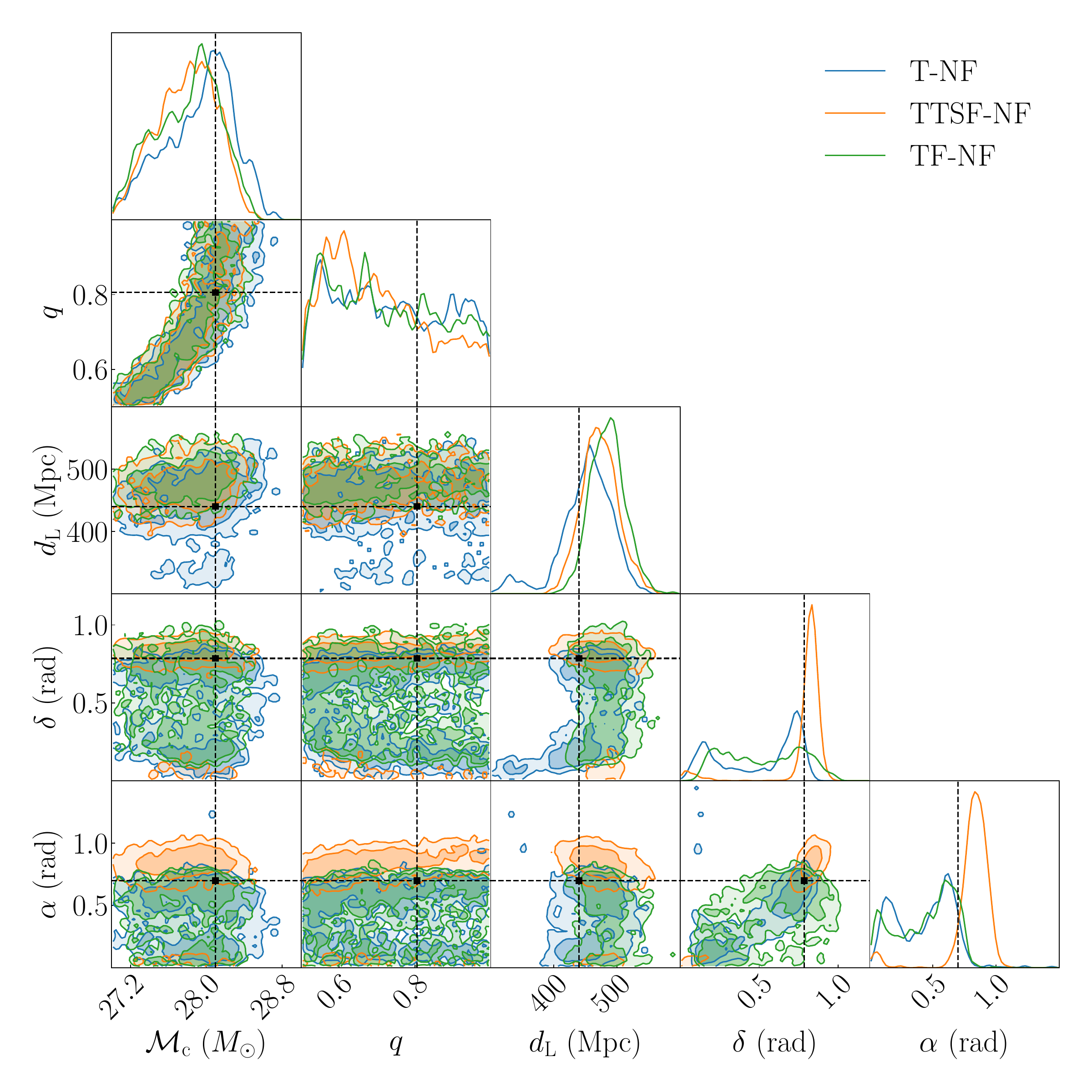}
\centering \caption{\label{fig8} One- and two-dimensional marginalization of $\mathcal{M}_{\rm c}$, $q$, $d_{\rm L}$, $\delta$ and $\alpha$ with blip for different NF estimates of the posterior distribution. The intersection points of the dashed lines are the injected parameters. Here, T-NF represents the posterior parameters predicted by the network using only time strain. TTSF-NF represents the posterior parameters predicted by the network using both time strain and a spectrogram. TF-NF represents the posterior parameters predicted by the network using only a spectrogram.}
\end{figure*}

Figure~\ref{fig8} presents the specific posterior distributions, notably in $\alpha$ and $\delta$. The time-spectrogram normalizing flow (TS-NF) reveals the widest posterior distribution, which is attributed to limitations in the time-frequency resolution and binning that may result in the loss of detailed information. Additionally, the posterior distribution is not significantly enhanced for T-NF. This is because the glitch's impact on the GW parameters primarily stems from temporal and frequency overlap. Consequently, TSF-NF struggles to differentiate between signal and noise information. TTSF-NF combines the strengths of both approaches. It emphasizes the contributions of different frequency components in the time series, facilitating a clearer distinction between signal and noise, while preserving time resolution as much as possible. This makes it more suitable for signal processing when the channel is contaminated by glitches. Despite the potential longer computation time required for spectrogram calculation using the $Q$-transform, the benefits of leveraging both representations outweigh the potential delay, particularly when considering the enhanced analytical capabilities they provide.


The results indicate that the posterior distributions of each NF for the luminosity distance $d_{\rm L}$, chirp mass $\mathcal{M}_{\rm c}$, and mass ratio $q$ are similar, which might be attributed to glitch extraction. However, it is premature to conclude that glitches affect different parameters inconsistently across neural networks. Table~\ref{tab4} presents the specific posterior distributions after data cleansing. Our findings reveal that the network performed best when utilizing both time strain data and spectrograms, achieving an accuracy comparable to that for glitch-free data. Notably, while the network inference's posterior distribution accuracy exceeds that of the Bilby inference, this discrepancy could stem from the non-stationary nature of noise at varying time instances.

\begin{table*}
\caption{A comparison between previously injected parameters and parameters recovered by {\tt Bilby} and normalizing flow methods. The recovered values are accompanied by their $2\sigma$ confidence regions. Here Bilby-offset represents the posterior parameters predicted by {\tt Bilby} with noise (without blip) plus GW signal. Bilby represents the posterior parameters predicted by {\tt Bilby} using noise (with blip) plus the GW signal.}\label{tab4}
\centering
\setlength\tabcolsep{12pt}
\renewcommand{\arraystretch}{1.5}
\begin{tabular}{ccccccc}
\hline \hline Parameter & Injected value & Bilby-offset & Bilby & T-NF & TTSF-NF & TS-NF \\
\hline $\mathcal{M}_{\rm c}~(M_{\odot})$ & 28.10 & $27.85_{-0.47}^{+0.28}$ & $28.28_{-0.53}^{+0.35}$& $27.97_{-0.86}^{+0.55}$& $27.79_{-0.66}^{+0.52}$& $27.55_{-0.76}^{+0.53}$\\
$q$ & 0.83 & $0.72_{-0.18}^{+0.26}$ & $0.67_{-0.15}^{+0.29}$& $0.72_{-0.21}^{+0.26}$& $0.69_{-0.18}^{+0.29}$& $0.71_{-0.20}^{+0.28}$\\
$d_{\rm L}~({\rm Mpc})$ & 440.00 & $423.57_{-38.89}^{+58.27}$ & $426.26_{-25.04}^{+27.59}$& $453.89_{-125.52}^{+64.94}$& $468.75_{-52.80}^{+60.76}$& $456.7_{-63.11}^{+63.11}$\\
$\alpha~({\rm rad})$ & 0.69 & $0.84_{-0.33}^{+0.15}$ & $2.89_{-0.16}^{+0.08}$& $0.46_{-0.42}^{+0.28}$& $0.83_{-0.16}^{+0.15}$& $0.36_{-0.58}^{+0.37}$\\
$\delta~({\rm rad})$ & 0.78 & $0.83_{-0.13}^{+0.03}$ & $-0.92_{-0.10}^{+0.32}$& $0.63_{-0.53}^{+0.17}$& $0.82_{-0.08}^{+0.07}$& $0.52_{-0.39}^{+0.40}$\\
\hline \hline
\end{tabular}
\end{table*}

The processing time for each network and the comparison with {\tt Bilby} are outlined in Table~\ref{tab5}. For the spectrogram-applied method, we considered the data processing time utilizing 32 CPUs in parallel. It is evident that regardless of the method used, the processing time is significantly faster than that of the traditional method. When combined with the $Q$-transform time, the overall processing time is not significantly longer, indicating the continued suitability of the $Q$-transform. In practical applications, additional time for calibration, strain data distribution, and signal identification must be considered. The time required for these tasks averages around 0.4 s for 2 s of time strain data \cite{George:2016hay,Viets:2017yvy}. All networks can complete processing within 2 s (the length of the data), making them suitable for real-time data processing.



\begin{table*}
\caption{A comparison of the computational time required by different methods for generating their respective samples. Note that the input data is noise (with blip) plus GW signal.}\label{tab5}
\centering
\setlength\tabcolsep{20pt}
\renewcommand{\arraystretch}{1.5}
\begin{tabular}{cccc}
\hline\hline
Sampling Method & Number of posterior samples & Total runtime (s) & Time per sample (s)\\
\hline
Bilby & 2111 & 246.41 & 0.1165 \\
T-NF & 1024 & 0.66 & 0.0006 \\
TTSF-NF & 1024 & 1.35 & 0.0013 \\
TS-NF & 1024 & 1.13 & 0.0011 \\
\hline\hline
\end{tabular}
\end{table*}

\subsection{Other glitches}

Given that various glitch types exhibit different distributions in time and frequency, where blips are typically short-term, it is crucial to demonstrate the algorithm's suitability for other glitch forms. To investigate this aspect, we chose scattered light noise for validation, aiming to determine if the method is effective for long-duration glitches. We utilized 1200 glitches for both training and testing.

\begin{figure}[!htp]
\centering


\includegraphics[width=0.5\textwidth]{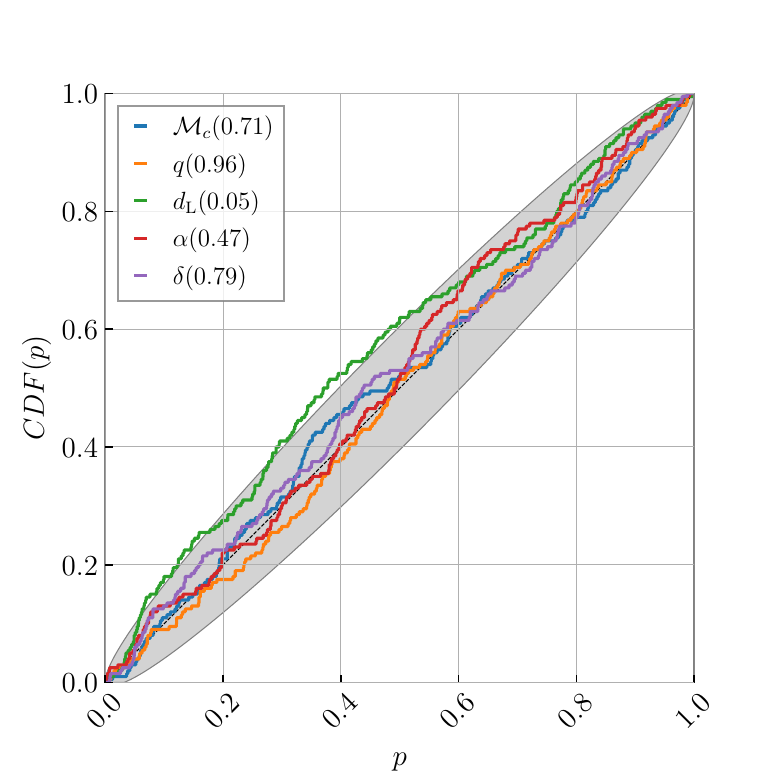}
\centering \caption{\label{fig9} P-P plot for 200 simulated datasets containing scattered light noises analyzed by TTSF-NF.}
\end{figure}

\begin{figure*}[!htp]
\centering
\includegraphics[width=0.5\textwidth]{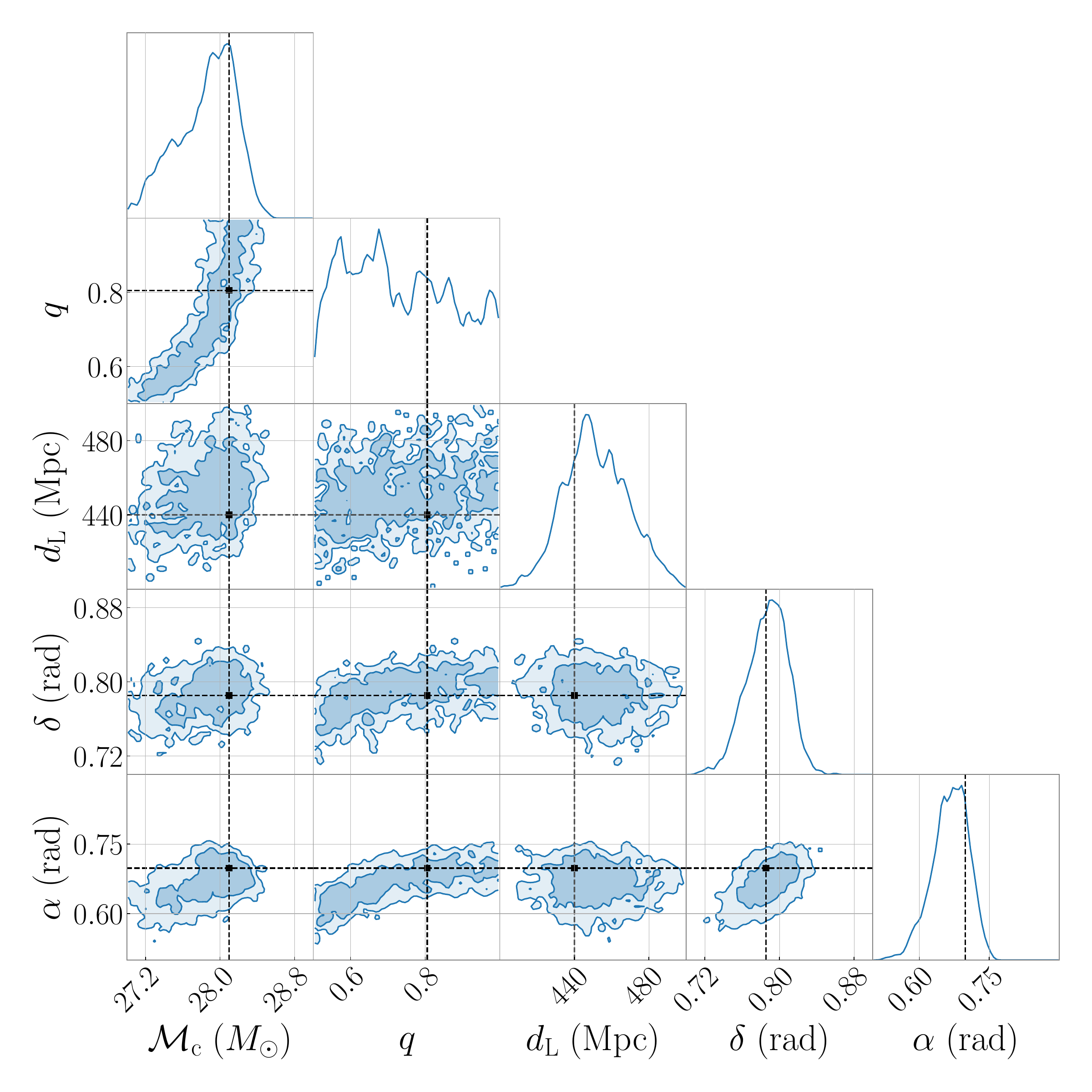}
\centering \caption{\label{fig10} One- and two-dimensional marginalized posterior distributions for $\mathcal{M}_{\rm c}$, $q$, $d_{\rm L}$, $\delta$ and $\alpha$ with scattered light noise. The intersection points of the dashed lines are the injected parameters.}
\end{figure*}

Figure~\ref{fig9} presents the P-P plot generated from 200 simulated datasets for testing. This plot underscores the reliability of the posterior distribution produced by our network, even when the noise includes scattered noise. Specific posterior distribution details are illustrated in Fig.~\ref{fig10}.



\section{Conclusion}\label{sec4}

GW detectors often encounter disruptive non-Gaussian noise artifacts known as glitches. The occurrence of these glitches in proximity to GW events can significantly impact subsequent parameter estimates. Conventional deglitching methods, while effective, demand substantial computational resources, posing challenges for achieving real-time deglitching at higher frequencies of GW events in the future. This, in turn, could impede timely observations of physical phenomena, including EM counterpart observations.

In this study, we leverage TTSF-NF to expedite parameter inference when GW data are marred by glitches. This assumes critical significance for swift wave source localization and real-time analysis, particularly as the frequency of future events escalates. Our pioneering approach involves combining the high temporal resolution of the time domain with the distinct discriminability of features in the time-frequency domain, aiming for rapid and judicious parameter inference in glitch-contaminated data. Notably, our choice of the normalizing flow as a more flexible flow contributes to the success of this innovative methodology.

Specifically, our focus was on glitches with a $\mathrm{SNR}>12$, which represents one of the most prevalent glitch types in GW detectors that existing robust methods struggle to effectively process. Notably, we discovered that relying solely on the spectrogram for parameter inference is suboptimal due to resolution limitations. Although the use of only the time strain on the training set produced effects equivalent to utilizing both time strain and a spectrogram, the network relying solely on time strain struggled to effectively discriminate between features in the time domain for previously unseen glitches. The incomplete separation of these features resulted in an inferior performance compared with the network utilizing both time strain and spectrogram. Our proposed method achieves real-time data processing, processing 2 s of data in 1.35 s. Additionally, we verified the applicability of this model to scattered light noise.

The integration of the normalizing flow opens promising avenues for the future real-time processing of glitch-contaminated data. It is essential to note that the upper limit of the frequency in this study is 1024 Hz, and as the frequency of BNS mergers exceeds this threshold, with BNS remaining in the detector's sensitive range for longer than 2 s, direct migration of this network to BNS scenarios is not viable. In our subsequent work, we plan to address the BNS scenario and {explore more suitable front-end network structures for better data fusion}. Simultaneously, we will consider implementing methods such as Cohen's Class of Time-Frequency Representations to enhance the time-frequency domain resolution for an optimal performance.

\begin{acknowledgments}
This research has made use of data or software obtained from the Gravitational Wave Open Science Center (gwosc.org), a service of LIGO Laboratory, the LIGO Scientific Collaboration, the Virgo Collaboration, and KAGRA.
We thank He Wang for helpful discussions. This work was supported by the National SKA Program of China (Grants Nos. 2022SKA0110200 and 2022SKA0110203), the National Natural Science Foundation of China (Grants Nos. 11975072, 11875102, and 11835009),  and the National 111 Project (Grant No. B16009).


\end{acknowledgments}

\bibliography{glitch_nf}

\end{document}